\def\hb{{\it HERA-B\/}}
\def\pt{$p^{}_T$}
\def\cp{$CP$\/}

\def\bpipi{$B^0\rightarrow\pi^+\pi^-$}
\def\bkpi{$B^0\rightarrow K^+\pi^-$}
\def\bdstarpi{$\overline{B}^{\,0}\rightarrow D^{*+}\pi^-$}
\def\bjpsiks{$B^0\rightarrow J/\psi K^0_S$}

\def\bsdspi{$B^0_s\rightarrow D^-_s\pi^+$}
\def\bsdslnu{$B^0_s\rightarrow D^-_s\ell^+\nu^{}_\ell$}
\def\bsdsppp{$B^0_s\rightarrow D^-_s\pi^+\pi^-\pi^+$}
\def\bskk{$B^0_s\rightarrow K^+K^-$}
\def\bspsiphi{$B^0_s\rightarrow J/\psi\phi$}

\documentstyle[11pt,epsfig]{article}
\begin{document}

\begin{titlepage}
\vspace*{0.5cm}
\begin{flushright}
{\large UCTP-119-99}
\end{flushright}
\vskip 0.15in
\large
\centerline {\bf \hb: Physics Potential and Prospects}
\normalsize
 
\vskip 2.0cm
\centerline {A.\ J.\ Schwartz~\footnote{schwartz@physics.uc.edu}}
\centerline {\it University of Cincinnati, Cincinnati, Ohio 45221}
\vskip 4.0cm
 
\centerline {\bf Abstract}
\vskip 1.0cm
\hb\ is a hadroproduction experiment located at DESY in Hamburg, Germany. 
The experiment produces $B$ mesons and baryons by inserting thin wire
targets into the halo of the proton beam circulating in the HERA storage
ring. The $B$ decays are studied to search for evidence of \cp\ violation 
and constrain the angle $\beta$ and possibly $\gamma$ and $\alpha$ of the 
CKM unitarity triangle. The experiment also produces $B_s$ mesons; these 
decays are studied to measure or constrain the mass 
difference $\Delta m^{}_s$ and width difference $\Delta\Gamma^{}_s$ 
between the two $B^0_s$/$\overline{B}_s^{\,0}$ mass eigenstates. 
Finally, the large number of $B$'s produced allows \hb\ to search for 
rare and forbidden decays such as $B\rightarrow K^{(*)}\ell^+_1\ell^-_2$. 
The experiment is scheduled to begin running in early 2000. 
\vfill
\end{titlepage}

\newpage
\section{Introduction}

\hb\ is a hadroproduction experiment that produces $b$ mesons
and baryons by colliding protons circulating in the HERA proton 
storage ring with fixed wire targets. The nominal interaction rate 
is 40~MHz. Downstream of the targets are the following detectors: 
a silicon-strip vertex detector used to reconstruct decay 
vertices; a large-aperture (250\,$\times$\,160 mrad$^2$) dipole 
magnet and thirteen drift chamber stations used to measure momentum;
a Ring-Imaging \u{C}erenkov counter (RICH) used to identify 
pions, kaons, and protons; an electromagnetic calorimeter used to 
measure electron and photon energies; and a set of gas ionization
chambers interleaved with iron plates used to identify muons. 
More details about these detectors can be found in the \hb\ 
Technical Design Report\,\cite{hbtdr}.

Two main triggers are used for the experiment: a dilepton trigger and a 
high-\pt\ trigger. The former identifies electron and muon pairs based on 
hits in the electromagnetic calorimeter or muon chambers. The hit positions
are used as seeds for a Kalman-filter tracking algorithm which searches 
regions in wire tracking chambers for corresponding hits. When hits 
in several successive chambers are found, a track is reconstructed. For 
events with two or more tracks, the invariant mass of all opposite-sign
track pairs ($m^{}_{\ell^+\ell^-}$) is calculated and events with 
$m^{}_{\ell^+\ell^-}$ above a threshold are passed. The threshold is 
chosen to efficiently select $J/\psi\rightarrow\ell^+\ell^-$ decays. 
The high-\pt\ trigger works in the same manner except that hits 
in a set of pad chambers located within the magnet determine the track 
seeds used by the Kalman filter. The hit patterns that initiate a trigger 
correspond to tracks that have high transverse momentum (\pt) with respect 
to the beam. The mass threshold used for the high-\pt\ trigger is chosen 
to efficiently select \bpipi, \bkpi, and \bskk\  
decays.\footnote{Charge-conjugate modes are included unless otherwise noted.}

Events that pass either the dilepton or high-\pt\ trigger are input to a 
second-level trigger running in a ``farm'' of dedicated processors. This 
trigger projects the two dilepton or high-\pt\ tracks into the silicon 
vertex detector, finds corresponding hits on silicon strips, and uses 
these hits to reconstruct silicon-based tracks. These silicon tracks 
are well-measured and are required to form a vertex some distance 
downstream of the nearest target wire.

To measure \cp\ violation, \hb\ measures the proper time distributions 
$d\Gamma(B^0\rightarrow f)/dt\/$ and 
$d\overline{\Gamma}(\overline{B}^{\,0}\rightarrow\overline{f})/dt\/$ 
and constructs the asymmetry parameter
${\cal A}^{}_{CP}(t) = (d\Gamma/dt - d\overline{\Gamma}/dt)/
(d\Gamma/dt + d\overline{\Gamma}/dt)$. 
For the final state $J/\psi K^0_S$, 
\begin{equation}
{\cal A}^{}_{CP}(t) \ =\ -D^{}_T D^{}_M\sin(2\beta)\sin(xt)\ ,
\label{eqn:cpasymmetry}
\end{equation}
where $D^{}_T$ is a ``dilution'' factor arising from mistagging
the $B^0$ or $\overline{B}^{\,0}$ flavor of the decay, $D^{}_M$ 
is a dilution factor caused by tagging signatures from neutral 
$B$ mesons which oscillate before they decay, $\beta$ 
is the interior angle of the CKM unitarity triangle, $x$ is the 
$B^0$-$\overline{B}^{\,0}$ mixing parameter $\Delta m/\Gamma$, and 
$t$ is the decay time in units of $B$ lifetime. Alternatively, \hb\ 
can count the total number of decays (corrected for acceptance) and 
construct the ratio of partial widths
${\cal A}^{}_{CP}=(\Gamma-\overline{\Gamma})/
					(\Gamma+\overline{\Gamma})$.
This quantity equals $-D^{}_T\,D^{}_M\,\sin(2\beta)\,\xi(t^{}_0)$,
where $\xi(t) = [\sin(xt) + x\cos(xt)]/(1+x^2)$ and $t^{}_0$ 
is the minimum value of decay time that a decay vertex must satisfy.
Such a requirement is necessary to sufficiently reduce backgrounds.
 
The proton interactions at \hb\ lead to a production asymmetry between 
$B^0$ and $\overline{B}^{\,0}$ of a few percent; this must be measured
and corrected for in order to detect a \cp\ asymmetry. The correction 
is made by constructing the ratios 
$$ R = \left(\frac{N^{}_{B\rightarrow f}}{\varepsilon^{}_f}\right)
\left(\frac{\varepsilon^{}_{n}}{N^{}_{B\rightarrow n}}\right)
\hspace*{0.25in}
{\rm and}
\hspace*{0.25in}
\overline{R} = \left(\frac{N^{}_{\overline{B}\rightarrow \overline{f}}}
{\varepsilon^{}_{\overline{f}}}\right)
\left(\frac{\varepsilon^{}_{\overline{n}}}
{N^{}_{\overline{B}\rightarrow \overline{n}}}\right), $$
where 
$N^{}_{B\rightarrow n}$ and $N^{}_{\overline{B}\rightarrow \overline{n}}$
are the numbers of $B^0$ and $\overline{B}^{\,0}$ decays to a 
(copious) channel {\it not\/} exhibiting \cp\ violation, e.g.,   
$B^0\rightarrow J/\psi\,K^{*0}\rightarrow\ell^+\ell^- K^+\pi^-$. 
The ratio $(R - \overline{R})/(R + \overline{R})$ then equals 
$(\Gamma - \overline{\Gamma})/(\Gamma + \overline{\Gamma})$.

For $N^{}_{B\rightarrow n} \gg N^{}_{B\rightarrow f}$, the 
uncertainty on $\sin(2\beta)$ is:
\begin{equation}
\Delta\sin(2\beta)\ \approx\  \frac{1}{\xi(t^{}_0)\,D^{}_M
\,D^{}_T}\ \frac{1}{\sqrt{\varepsilon^{}_{\rm tag}\,N^{}_B}}\ 
\sqrt{\frac{S+B}{S}}\ ,
\label{eqn:err_timeint}
\end{equation}
where $N^{}_B$ is the number of \bjpsiks\ candidates,
$S$ is the number of true signal events, 
$B$ is the number of background events, and
$\varepsilon^{}_{\rm tag}$ is the tagging efficiency. The factor 
$D^{}_T\sqrt{\varepsilon^{}_{\rm tag}}$ indicates the performance 
of the tagging method. There are four methods to be used in \hb\ to 
tag the flavor of the decaying $B^0$ or $\overline{B}^{\,0}$:
\begin{enumerate}
\item the charge of a lepton with $p\ (p^{}_T)\,>\,5\ (0.8)$~GeV/$c$
that originates from the semileptonic decay of the other $b$ hadron 
in the event;
\item the charge of a kaon with $5<p<50$~GeV/$c$ that originates from
the $b\rightarrow D\rightarrow K^\pm$ decay of the other $b$ hadron 
in the event;
\item the ``jet charge'' defined by:
\begin{equation}
Q^{}_{\rm jet}\ =\ {\displaystyle \sum_i}\,q^{}_i 
	\left| {\bf \vec{p}}^{}_i\cdot{\bf \hat{a}}\right|
\ /\ 
{\displaystyle \sum_i}\,\left| {\bf \vec{p}}^{}_i\cdot{\bf \hat{a}}\right|\ ,
\end{equation}
where ${\bf \hat{a}}$ is the direction vector of the $B$ candidate as 
defined by the positions of the interaction and decay vertices, 
and $i$ runs over all tracks in the decay vertex;
\item the ``same-side'' tag, which is the charge of a track within a small
$\eta$-$\phi$ cone around the $B^0$ momentum vector which has minimum \pt\ 
relative to the combined $(B^0\,+\,{\rm track})$ momentum vector.
\end{enumerate}
The first three methods have been studied extensively using a Monte Carlo 
simulation, and their expected efficiencies, dilution factors, and overall 
factors $D^{}_MD^{}_T\sqrt{\varepsilon^{}_{\rm tag}}$ are listed in 
Table~\ref{tab:effic}. 

\begin{table}[htb]
\begin{center}
\begin{tabular}{|l|ccc|}
\hline
{\bf Method} & {\bf {\boldmath $\varepsilon^{}_{\rm tag}$}} & 
{\bf {\boldmath $D^{}_M D^{}_T$}} & 
{\bf {\boldmath $D^{}_M D^{}_T\sqrt{\varepsilon^{}_{\rm tag}}$}} \\
\hline
$p^{}_\ell\ (p^{}_T)\,>\,5\ (0.8)$~GeV/$c$  &  0.15 & 0.53 & 0.20 \\   
	&	&	&	\\
$5<p^{}_K<50$~GeV/$c$ & 0.48 & 0.29 & 0.20 \\   
	&	&	&	\\
${\displaystyle \sum_i}\,q^{}_i \left| {\bf \vec{p}}^{}_i\cdot{\bf \hat{a}}\right|
\ {\displaystyle /}\ 
{\displaystyle \sum_i}\,\left| {\bf \vec{p}}^{}_i\cdot{\bf \hat{a}}\right|$
	& 0.99 & 0.12 & 0.12 \\   \hline
{\bf All combined:} &   &   &  {\bf 0.29} \\   
\hline
\end{tabular}
\end{center}
\caption{The expected efficiencies, dilution factors $D^{}_M D^{}_T$, 
and overall factors $D^{}_MD^{}_T\sqrt{\varepsilon^{}_{\rm tag}}$  
for three tagging methods to be used in \hb.}
\label{tab:effic}
\end{table}

\section{\bf {\boldmath Measurement of $\sin(2\beta)$ and $\sin(2\alpha)$}}

To estimate the precision with which \hb\ can measure $\sin(2\beta)$
using \bjpsiks\ decays, we insert values for all parameters into 
Eq.~(\ref{eqn:err_timeint}). These values are listed in 
Table~\ref{tab:sin2beta}; the resultant uncertainty is 
$\Delta\sin(2\beta)=0.17$ in one year of running. The precision with 
which \hb\ can measure $\sin(2\alpha)$ using \bpipi\ decays is also 
estimated using Eq.~(\ref{eqn:err_timeint}); inserting values for
the parameters (see Table~\ref{tab:sin2alpha}) gives
$\Delta\sin(2\alpha)=0.36$ in three years of running. 
This estimate does not include theoretical uncertainty, which is
discussed below. The branching fraction for \bpipi\ is taken to be
$0.47\times 10^{-5}$\,\cite{cleo_pipi}, and the signal-to-background 
ratio is taken to be one. Both estimates above include a lifetime cut on 
the decay vertex of $0.5\tau^{}_B$, which is approximately 10 times 
the expected resolution in the vertex $z$ position. The precision obtained 
for $\sin(2\beta)$ would be a significant improvement over the current 
result from CDF\,\cite{cdfsin2beta}; the precision obtained for 
$\sin(2\alpha)$ would constrain this parameter to a range much 
smaller than that allowed by current data\,\cite{alphaprec}.

\begin{table}[htb]
\begin{center}
\begin{tabular}{|l|c|}
\hline
$\sigma_{b\bar{b}}$         & 12\,nb       \\
\hline
$\sigma_{inel}$             & 13\,mb       \\
$\sigma_{b\bar{b}}/\sigma_{inel}$    & $9.2\cdot 10^{-7}$    \\ 
interaction rate          & 40\,MHz      \\
$b\bar{b}$ rate             & 37\,Hz       \\
fraction reconstructed as $B^0\rightarrow J/\psi K^0_S$ 
					&  $3.8 \cdot 10^{-6}$ \\
\ \ \ (for $f(b\bar{b}\rightarrow B^0X)\simeq 0.7$) 
           &      \\
reconstructed $B^0\rightarrow J/\psi K^0_S$\ \,per $10^7$\,s        
                                       &   1400 \\
statistical factor $1/\xi(\tau^{}_B)$  &  1.5  \\
$\langle 1/D^{}_M\rangle$ mixing of tagging $B$   &  1.2  \\
$\ell + K + \sum q^{}_i$\ \,tagging factor 
$(D^{}_T\sqrt{\epsilon^{}_{\rm tag}})^{-1}$	&   3.5   \\
\hline
{\boldmath\bf $\Delta\sin(2\beta)$ after $10^7$\,s} & {\bf 0.17}   \\
\hline
\end{tabular}
\end{center}
\caption{The expected uncertainty in $\sin(2\beta)$ after one year 
of running.}
\label{tab:sin2beta}
\end{table}

\begin{table}[htb]
\begin{center}
\begin{tabular}{|l|c|}
\hline
$\sigma_{b\bar{b}}$         & 12\,nb       \\
\hline
$\sigma_{inel}$             & 13\,mb       \\
$\sigma_{b\bar{b}}/\sigma_{inel}$    & $9.2\cdot 10^{-7}$    \\ 
interaction rate          & 40\,MHz      \\
$b\bar{b}$ rate             & 37\,Hz       \\
fraction reconstructed as $B^0\rightarrow\pi^+\pi^-$ 
        			   &      $5.6\cdot 10^{-7}$  \\
\ \ \ (for $f(b\bar{b}\rightarrow B^0X)\simeq 0.7$ and
		$B^{}_{B\rightarrow\pi\pi}=0.47\times 10^{-5}$) 
& \\
$B^0\rightarrow\pi^+\pi^-$ decays per $10^7$\,s        
                                       &   210  \\
statistical factor $1/\xi(\tau^{}_B)$  &  1.5  \\
$\langle 1/D^{}_M\rangle$ mixing of tagging $B$   &  1.2  \\
$\ell + K + \sum q^{}_i$\ \,tagging factor 
$(D^{}_T\sqrt{\epsilon^{}_{\rm tag}})^{-1}$	&   3.5   \\
\hline
{\boldmath\bf $\Delta\sin(2\alpha)$ after $3\times 10^7$\,s}
                            & {\bf 0.25}   \\
(assuming $S/B\approx 1$)   &  (0.36)    \\
\hline
\end{tabular}
\end{center}
\caption{The expected uncertainty in $\sin(2\alpha)$ after three years 
of running. For simplicity, we assume the ratio of penguin amplitude 
to tree amplitude to be very small.}
\label{tab:sin2alpha}
\end{table}

In addition to \bjpsiks\ and \bpipi\ decays, there are other modes
which provide sensitivity to $\sin(2\beta)$ and $\sin(2\alpha)$. These 
are summarized in Tables~\ref{tab:beta} and \ref{tab:alpha}.
Each table has six columns:
the left-most column lists the decay mode of interest; the second column 
lists the branching fraction for the decay as either measured from data
or estimated in Ref.~\cite{Babar_book}. 
The third column lists the number of $B$\,+\,$\overline{B}$ decays 
reconstructed per year. This number is obtained by multiplying the  
branching fraction by the $b\bar{b}$ production rate, the fraction 
of $b\bar{b}$ pairs which yield a $B$ meson, the factor $10^7$ seconds 
of running per year, and the estimated \hb\ trigger and reconstruction 
efficiencies.
The fourth column lists the first-level trigger which accepts the 
decay. The fifth column lists the number of vertices arising from 
the decay chain and the number of kinematic constraints that can be 
used to reject background. The sixth column shows a sketch of the decay 
topology. These last two columns indicate the level of background 
suppression one expects for the decay mode: two or more decay vertices 
and a large number of kinematic constraints should provide substantial 
background suppression. 

Experimentally, the modes listed in Table~\ref{tab:beta} are very 
promising: all have two decay vertices and at least five kinematic 
contraints. The branching fractions are substantial and indicate that 
\hb\ should reconstruct $>$\,3000 of these decays per year. Theoretically, 
the decays provide a clean measurement of $\sin(2\beta)$: the dominant 
amplitudes have a single weak phase ($\mbox{Arg}\,V^*_{cs}V^{}_{cb}$), 
and thus {\it direct\/} \cp\ violation is negligible and 
Eq.~(\ref{eqn:cpasymmetry}) has negligible theoretical uncertainty.

The decay modes listed in Table~\ref{tab:alpha} are more challenging: 
they have only one decay vertex and fewer kinematic contraints.
Theoretically, the amplitudes may receive significant contributions 
from penguin diagrams which have a different weak phase from that of 
the tree diagram ($\mbox{Arg}\,V^*_{cd}V^{}_{cb}$); in this case
Eq.~(\ref{eqn:cpasymmetry}) also has a $\cos (xt)$ dependence and,
in addition, the measured value of $\alpha$ is shifted from the true 
value by an amount which depends on the unknown ratio of the penguin 
amplitude to the tree amplitude. This shift can be determined via
an isospin analysis\,\cite{London}, but such an analysis requires 
measuring small branching fractions of difficult-to-reconstruct 
decay modes such as $B^+\rightarrow \pi^+ \pi^0$ and 
$B^0\rightarrow\pi^0\pi^0$. Another method\,\cite{Rosner} to determine 
$\alpha$ in the presence of penguins uses $SU(3)$ flavor symmetry and 
the decays \bpipi, \bkpi, and $B^+\rightarrow\pi^+K^0$. Although the 
last decay will not be measured in \hb\ (it is not selected by the 
trigger), only the branching fraction is needed and this is expected 
to be well-measured at CLEO~III\,\cite{CLEOIII}, 
{\it BaBar\/}\,\cite{Babar_book,Babar}, and {\it BELLE\/}\,\cite{BELLE}.  
The three-body decays $B^0\rightarrow\rho\pi\rightarrow\pi^+\pi^-\pi^0$
can also determine $\alpha$ in the presence of penguins via a
Dalitz plot analysis\,\cite{Quinn}; however, the branching fraction 
for $B^0\rightarrow\rho^0\pi^0$ is probably too low for this method 
to be useful in \hb. The modes $B^0\rightarrow\rho\rho$ which have 
two vector particles in the final state require an angular analysis 
to extract $\sin(2\alpha)$, and this in turn requires a relatively 
large event sample.

\section{\bf {\boldmath Measurement of $\sin(2\gamma)$}}

A number of decay modes have been discussed in the literature which 
provide sensitivity to the angle $\gamma$, which is the third angle 
of the CKM unitarity triangle. 
Some of these  modes are listed in Table~\ref{tab:gamma}. The first three 
were proposed by Gronau and Wyler\,\cite{GronauWyler}; by measuring their 
branching fractions and those of their charge conjugates, one can construct 
two triangles with a common base. The angle between the sides representing 
the $B^+\rightarrow D^0_{CP}K^+$ and $B^-\rightarrow D^0_{CP}K^-$ amplitudes 
is $2\gamma$. Because only tree diagrams contribute to these decays, there 
is no uncertainty arising from penguin diagrams or final-state 
rescattering.\footnote{However, if the $D^0$ is detected via a hadronic 
decay such as $D^0\rightarrow K^-\pi^+$, there is uncertainty arising 
from interference with a doubly-Cabibbo-suppressed amplitude such as 
$\overline{D}^{\,0}\rightarrow K^-\pi^+$ --\,see Ref.~\cite{Soni}.}
Unfortunately, the expected branching fraction for 
$B^-\rightarrow\overline{D}^{\,0}K^-$ is too low for \hb\ to make a 
meaningful measurement. The next three modes were proposed by 
Dunietz\,\cite{Dunietz} and are analogous to the first three but 
pertain to $B^0$ decays; their branching fractions are also 
expected to be too low for \hb.

The amplitudes for $B^0\rightarrow K^+\pi^-$ and 
$\overline{B}^{\,0}\rightarrow K^-\pi^+$ can be used in conjunction with 
that for $B^+\rightarrow\pi^+ K^0$ (measured at CLEO/{\it BaBar}/{\it BELLE})
to construct two more triangles and determine $\sin(2\gamma)$. This 
method\,\cite{Fleischer} requires estimating a color-allowed tree amplitude, 
and this introduces theoretical uncertainty. However, the branching fraction 
for $B^0\rightarrow K^+\pi^-$ is large enough for \hb\ to collect a few 
hundred events per year, and such a sample could provide a 
first constraint on this angle. It is estimated\,\cite{Danilov} that in 
one year of running, \hb\ could measure a difference in the rates for 
$B^0\rightarrow K^+\pi^-$ and $\overline{B}^{\,0}\rightarrow K^-\pi^+$ 
with a precision $\Delta {\cal A}^{}_{CP}/{\cal A}^{}_{CP} = 0.08\ (0.16)$,
for a signal-to-background ratio of unity~(0.10). 

A bound $\gamma^{}_0$ of the form $0<\gamma <\gamma^{}_0$ or 
$\gamma^{}_0 < \gamma < 180^\circ$ which may have less theoretical 
uncertainty\,\cite{FleischerMannel} can be obtained from the inclusive 
measurement 
$B(\overline{B}^{\,0}\rightarrow K^-\pi^+ + B^0\rightarrow K^+\pi^-)$; 
however, this bound depends on the ratio 
$B(\overline{B}^{\,0}\rightarrow K^-\pi^+ + B^0\rightarrow K^+\pi^-)/
B(B^+\rightarrow K^0\pi^+ + B^-\rightarrow \overline{K}^{\,0}\pi^-)$
being less than one, which may not be the case 
(e.g., see Ref.~\cite{cleo_pipi}).

The decay \bdstarpi\ is also promising, as the branching fraction is 
substantial and 2000--3000 of these decays should be reconstructed per 
year. Because the final state is not a \cp\ eigenstate, fitting for the 
decay time dependence yields $\sin(2\beta + \gamma)$ rather than 
$\sin (2\gamma)$, which is advantageous if $\gamma$ is near $90^\circ$ 
(as possibly indicated by data from CLEO\,\cite{cleo_pipi,Wurthwein}). 
This decay proceeds via the $b\rightarrow c\bar{u}d$ transition; 
because all quark flavors 
in the final state are different, only tree diagrams contribute 
and there is no theoretical uncertainty arising from penguin diagrams 
or final-state rescattering. The ultimate success of this method and 
the others discussed above depends on as-yet-unknown backgrounds.

\section{\bf {\boldmath $B^0_s$ Physics}}

In addition to $B^{}_d$ mesons, \hb\ also produces $B^{}_s$ mesons; 
these allow the experiment to measure or constrain parameters of the 
$B^0_s$-$\overline{B}^{\,0}_s$ system. For example, the decays \bsdslnu, 
\bsdspi, \bsdsppp, $B^0_s\rightarrow J/\psi\,\overline{K}^{\,*0}$, 
and \bskk\ can be used to measure the mass difference 
$\Delta m^{}_s$ or decay width difference $\Delta\Gamma^{}_s$ between 
the two $B^0_s$/$\overline{B}^{\,0}_s$ mass eigenstates. The mode
$B^0_s\rightarrow D^-_s K^+$
(analogous to \bdstarpi, i.e., no penguin contribution) can be used
to measure $2\phi^{}_M + \gamma$, where $\phi^{}_M$ is the 
$B^0_s$-$\overline{B}^{\,0}_s$ mixing phase (expected to be very small). 
The mode \bspsiphi\ should not exhibit \cp\ violation at \hb's level 
of sensitivity\,\cite{Buraspsiphi}, and thus observing a $CP$ asymmetry 
would indicate physics beyond the Standard Model. For all these modes, 
the number of events \hb\ expects to reconstruct per year is listed in 
Table~\ref{tab:bsphysics}. For most modes, the number of reconstructed
events is $>$\,100 per year, or $>$\,300 events in three years.
Unfortunately, the number of reconstructed 
$B^0_s\rightarrow D^-_s K^+$ decays is too low for \hb\ to constrain 
$2\phi^{}_M + \gamma$; this sample needs to be large in order to 
resolve the rapid $B^0_s$-$\overline{B}^{\,0}_s$ oscillations.

The ``reach'' of \hb\ for measuring the mass difference $\Delta m^{}_s$ 
has been estimated based on the amplitude fit method of Ref.~\cite{Moser}. 
This reach is taken to be that value of $\Delta m^{}_s$ in which the 
quantity $S/N$ (signal to noise ratio) equals 1.645, where $S/N$ is
given by:
\begin{equation}
\frac{S}{N}\ =\ 
\sqrt{\frac{n}{2}}\ f^{}_s\ (1-2\eta ) e^{-(\sigma^{}_t\,\Delta m )^2/2}.
\end{equation}
In this expression, $n$ is the number of $B^0_s$ or $\overline{B}^{\,0}_s$
decays reconstructed, $f^{}_s$ is the fraction which are true signal, 
$\eta$ is the probability of a mistag, and $\sigma^{}_t$ is the decay 
time resolution. Typical values for these parameters are listed in 
Table~\ref{tab:bsmixing} and imply that in three years of running, 
$\Delta m^{}_s$ could be measured or constrained up to values 
$>$\,24~ps$^{-1}$. This is superior to the current world limit  
$\Delta m^{}_s > 14.3$ (95\% C.L.)\,\cite{xslimit}.

\hb\ can search for \cp\ violation in the $B^0_s$ system by measuring 
the decay time distribution of \bskk\ decays and looking for a deviation 
from an exponential time dependence.\footnote{A non-exponential decay time 
distribution demonstrates interference between the $B_{s, H}$ and $B_{s, L}$ 
amplitudes to $K^+K^-$; thus there exists an interference term in the 
rate $|\langle K^+K^- |H|B^0\rangle |^2$ which changes sign under 
$CP$. Such a term is $CP$ violating --\,see e.g., Ref.~\cite{Sachs}.}
If the time distribution {\it is\/} exponential, then it can be used 
along with that for $B^0_s\rightarrow D^-_s\pi^+$ to determine 
the width difference $\Delta\Gamma^{}_s$\,\cite{Liu}. Since 
$dN^{}_{KK}/dt \propto\ e^{-\Gamma^{}_1t}\doteq e^{-\Gamma^{}_L t}$, and 
$dN^{}_{D^{}_s\pi}/dt \propto\ e^{-(\Gamma^{}_1+\Gamma^{}_2)t/2}
\cosh (\Delta\Gamma^{}_s/2)t \approx e^{-(\Gamma^{}_1+\Gamma^{}_2)t/2}
\doteq e^{-(\Gamma^{}_L+\Gamma^{}_H)t/2}$,
$\Gamma^{}_{D^{}_s\pi}-\Gamma^{}_{KK} = 
(\Gamma^{}_L+\Gamma^{}_H)/2 - \Gamma^{}_L = \Delta\Gamma^{}_s/2$.
\hb\ expects to reconstruct $\sim$\,35 \bskk\ decays per year, 
and a preliminary study indicates that such a sample could be
sensitive to a difference $\Delta\Gamma^{}_s/\Gamma^{}_s\approx 0.18$. 
This sensitivity is superior to that of the current measurement from 
CDF\,\cite{cdfwidthdiff}.

\section{Summary}
In summary, the physics potential of \hb\ is substantial. 
We expect the experiment to do well measuring $\sin(2\beta)$, 
with an expected uncertainty of $\Delta\sin(2\beta)=0.17$ after
one year of running. The experiment can constrain $\sin(2\alpha)$,
but this requires a better theoretical understanding of the penguin
contribution to this mode. If the penguin contribution were very small, 
\hb\ could obtain $\Delta\sin(2\alpha)\approx 0.36$ after three years 
of running for a signal-to-background ratio of unity. The experiment can 
potentially constrain the angle $\gamma$ via \bkpi\ and \bdstarpi\ decays; 
the viability of these methods depends on unknown backgrounds. \hb\ can 
constrain $\Delta m^{}_s$ to be $>$\,24~ps$^{-1}$ in three years of running, 
which is an improvement over the current world limit of $\Delta m^{}_s > 14.3$ 
(95\% C.L.). \hb\ will also search for rare and forbidden dilepton decays 
such as $B\rightarrow K^*\ell^+\ell^-$, $B^0_s\rightarrow \phi\ell^+\ell^-$, 
and $B\rightarrow K^{(*)}\mu^\pm e^\mp$\ \cite{AliBall}. Finally, \hb\ may 
find new phenomena such as measurable \cp\ violation in \bspsiphi\ decays.


\begin{table}[htb]
\begin{center}
\hbox{
\hspace*{-0.40in}
\begin{tabular}{|c|ccccc|}
\hline
{\bf Decay mode} &  {\bf BR} &  {\bf Rec.\,Evnts} & {\bf Trigger}  
& {\bf Vertices/}{\bf }  & {\bf Topology} \\
     &  {\bf ($\times 10^{-5}$)} & {\bf per year}  &    &
						{\bf Constraints}  &   \\
\hline
  &   &   &  &  &  \\
{\bf {\boldmath $B^0\rightarrow J/\psi\,K^0_S$}}
	& 3.7  & 1400 & $\mu\mu/ee$  & 2/5 & 
		\mbox{\epsfig{file=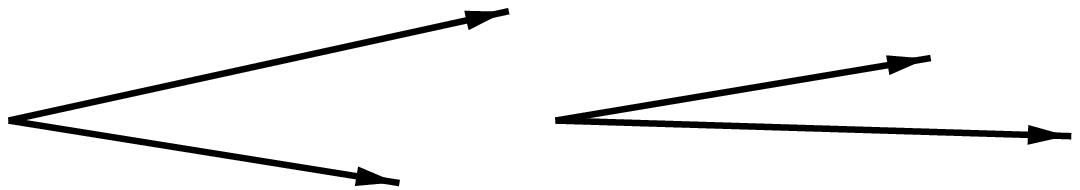,height=0.22in}} \\
$(J/\psi\rightarrow\ell^+\ell^-)$  &   &   &  &  &  \\
$(K^0_S\rightarrow\pi^+\pi^-)$  &   &   &  &  &  \\
  &   &   &  &  &  \\
{\bf {\boldmath $B^0\rightarrow J/\psi\,K^{*0}$}}
	& 1.9  & 700  & $\mu\mu/ee$  & 2/7  & 
		\mbox{\epsfig{file=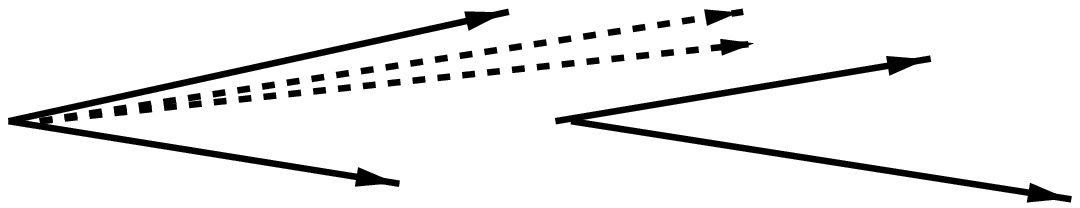,height=0.25in}} \\
$(J/\psi\rightarrow\ell^+\ell^-)$  &   &   &  &  &  \\
$(K^{*0}\rightarrow K^0_S\pi^0)$  &   &   &  &  &  \\
  &   &   &  &  &  \\
{\bf {\boldmath $B^0\rightarrow \psi'\,K^0_S$}}
	& 0.87   & 330   & $\mu\mu/ee$  & 2/6  & 
		\mbox{\epsfig{file=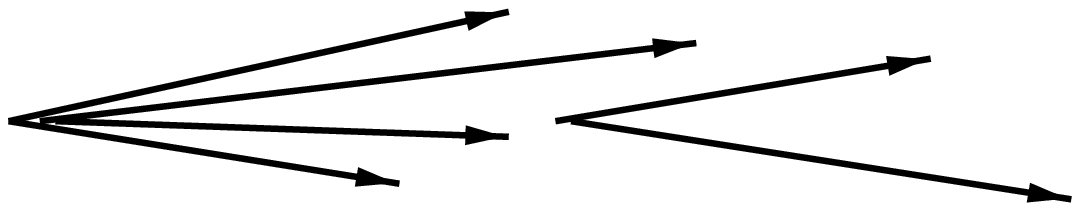,height=0.25in}} \\
$(\psi'\rightarrow J/\psi\,\pi^+\pi^-)$  &   &   &  &  &  \\
$(J/\psi\rightarrow\ell^+\ell^-)$  &   &   &  &  &  \\
  &   &   &  &  &  \\
{\bf {\boldmath $B^0\rightarrow \psi'\,K^{*0}$}}
	& 0.58  & 220  & $\mu\mu/ee$  & 2/8  & 
		\mbox{\epsfig{file=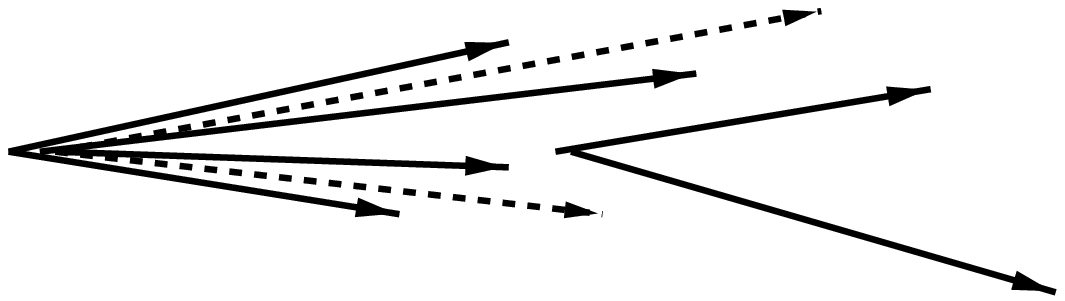,height=0.35in}} \\
$(\psi'\rightarrow J/\psi\,\pi^+\pi^-)$  &   &   &  &  &  \\
$(J/\psi\rightarrow\ell^+\ell^-)$  &   &   &  &  &  \\
$(K^{*0}\rightarrow K^0_S\pi^0)$  &   &   &  &  &  \\
  &   &   &  &  &  \\
{\bf {\boldmath $B^0\rightarrow \chi^{}_{c1}\,K^0_S$}}
	& 1.1  & 430  & $\mu\mu/ee$  & 2/6  &  
		\mbox{\epsfig{file=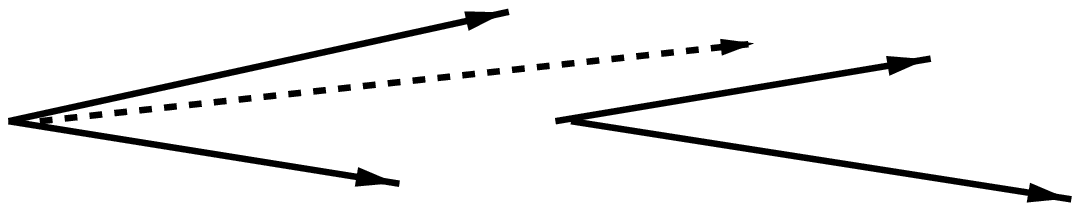,height=0.25in}} \\
$(\chi^{}_{c1}\rightarrow J/\psi\,\gamma)$  &   &   &  &  &  \\
$(J/\psi\rightarrow\ell^+\ell^-)$  &   &   &  &  &  \\
  &   &   &  &  &  \\
{\bf {\boldmath $B^0\rightarrow \chi^{}_{c1}\,K^{*0}$}}
	& 0.52  & 200  & $\mu\mu/ee$  & 2/8  & 
		\mbox{\epsfig{file=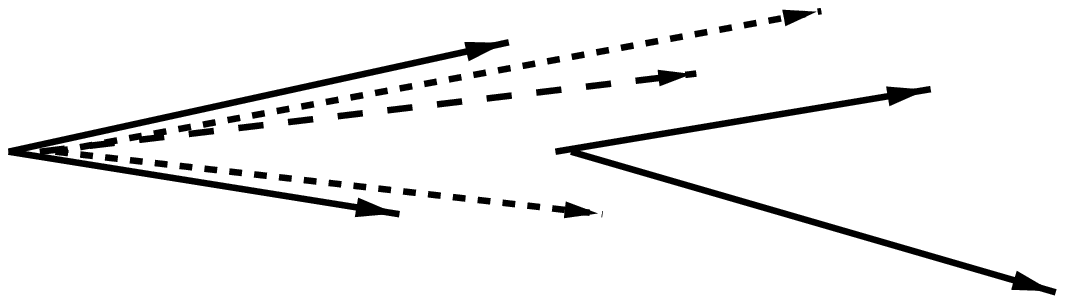,height=0.35in}} \\
$(\chi^{}_{c1}\rightarrow J/\psi\,\gamma)$  &   &   &  &  &  \\
$(J/\psi\rightarrow\ell^+\ell^-)$  &   &   &  &  &  \\
$(K^{*0}\rightarrow K^0_S\pi^0)$  &   &   &  &  &  \\
  &   &   &  &  &  \\
\hline
\end{tabular}
}
\end{center}
\caption{Methods to measure the CKM angle $\beta$ in \hb.}
\label{tab:beta}
\end{table}

\begin{table}[htb]
\begin{center}
\hbox{
\hspace*{-0.30in}
\begin{tabular}{|l|ccccc|}
\hline
{\bf Decay mode} &  {\bf BR} &  {\bf Rec.\,Evnts} & {\bf Trigger}  
& {\bf Vertices/}{\bf }  & {\bf Topology} \\
     &  {\bf ($\times 10^{-5}$)} & {\bf per year}  &    &
						{\bf Constraints}  &   \\
\hline
  &   &   &  &  &  \\
{\bf {\boldmath $B^0\rightarrow \pi^+\pi^-$}}
	& 0.47  & 210 & high-$p^{}_T$  & 1/2 & 
		\mbox{\epsfig{file=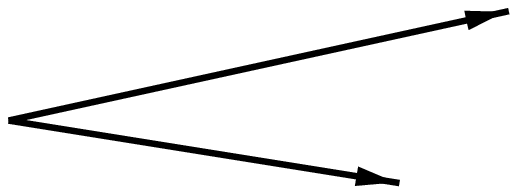,height=0.25in}} \\
{\bf {\boldmath $B^0\rightarrow \pi^- K^+$}}
	& 1.4  & 260 & high-$p^{}_T$  & 1/2 & 
		\mbox{\epsfig{file=topG.eps,height=0.25in}} \\
\hline
  &   &   &  &  &  \\
{\bf {\boldmath $B^0\rightarrow \rho^+\pi^-$}}
	& 4.4  & $<1900$  & high-$p^{}_T$  & 1/4  & 
		\mbox{\epsfig{file=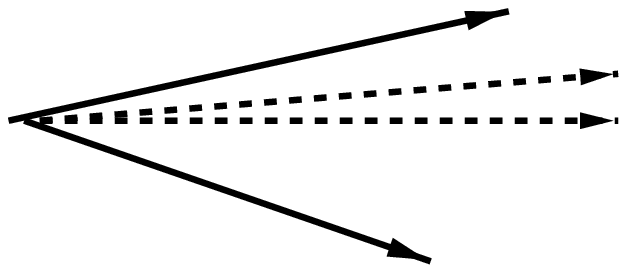,height=0.32in}} \\
{\bf {\boldmath $B^0\rightarrow \rho^-\pi^+$}}
	& 1.0  & $<420$  & high-$p^{}_T$  & 1/4  & 
		\mbox{\epsfig{file=topH.eps,height=0.32in}} \\
{\bf {\boldmath $B^0\rightarrow \rho^0\pi^0$}}
	& 0.1  & $<42$  & high-$p^{}_T$  & 1/4  & 
		\mbox{\epsfig{file=topH.eps,height=0.32in}} \\
\hline
  &   &   &  &  &  \\
{\bf {\boldmath $B^0\rightarrow \rho^+\rho^-$}}
	& 2.9   & $\ll 1200$   & high-$p^{}_T$  & 1/6  & 
		\mbox{\epsfig{file=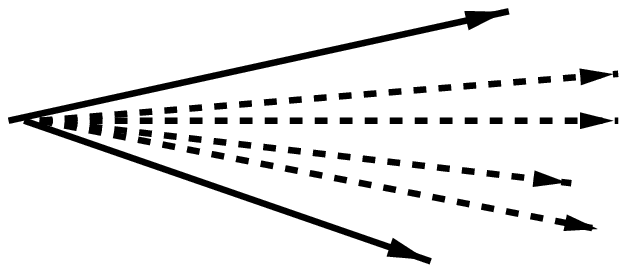,height=0.32in}} \\
{\bf {\boldmath $B^0\rightarrow \rho^0\rho^0$}}
	& 0.064   & $\ll 27$   & high-$p^{}_T$  & 1/4  & 
		\mbox{\epsfig{file=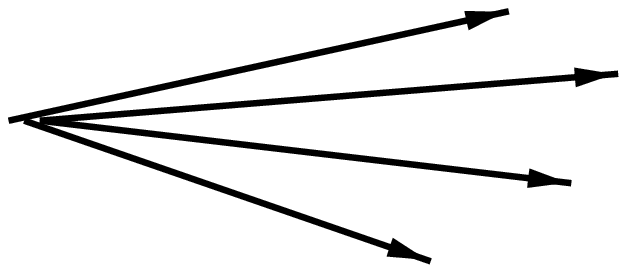,height=0.32in}} \\
  &   &   &  &  &  \\
\hline
\end{tabular}
}
\end{center}
\caption{Methods to measure the CKM angle $\alpha$ in \hb.
A ``$<$'' sign indicates that the trigger and reconstruction 
efficiencies assumed are optimistic.}
\label{tab:alpha}
\end{table}

\begin{table}[htb]
\begin{center}
\hbox{
\hspace*{-0.40in}
\begin{tabular}{|l|ccccc|}
\hline
{\bf Decay mode} &  {\bf BR} &  {\bf Rec.\,Evnts} & {\bf Trigger}  
& {\bf Vertices/}{\bf }  & {\bf Topology} \\
     &  {\bf ($\times 10^{-5}$)} & {\bf per year}  &    &
						{\bf Constraints}  &   \\
\hline
  &   &   &  &  &  \\
{\bf {\boldmath $B^-\rightarrow D^0 K^-$}}
	& 1.5  & $<650$ & high-$p^{}_T$  & 1/4 & 
		\mbox{\epsfig{file=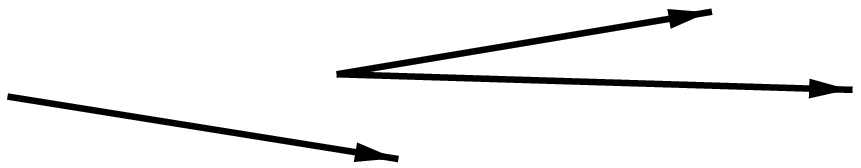,height=0.22in}} \\
$(D^0\rightarrow K^-\pi^+)$  &   &   &  &  &  \\
  &   &   &  &  &  \\
{\bf {\boldmath $B^-\rightarrow\overline{D}^0 K^-$}}
	& 0.0077  & $<3$ & high-$p^{}_T$  & 1/4 & 
		\mbox{\epsfig{file=topAx.eps,height=0.22in}} \\
$(\overline{D}^0\rightarrow K^+\pi^-)$  &   &   &  &  &  \\
  &   &   &  &  &  \\
{\bf {\boldmath $B^-\rightarrow D^0_{CP} K^-$}}
	& 0.23  & $<98$ & high-$p^{}_T$  & 1/4 & 
		\mbox{\epsfig{file=topAx.eps,height=0.22in}} \\
$(D^0_{CP}\rightarrow K^+ K^-)$  &   &   &  &  &  \\
$(D^0_{CP}\rightarrow\pi^+\pi^-)$  &   &   &  &  &  \\
\hline
  &   &   &  &  &  \\
{\bf {\boldmath $B^0\rightarrow \overline{D}^0 K^{*0}$}}
	&  0.023  & $<10$ & high-$p^{}_T$  & 2/5 & 
		\mbox{\epsfig{file=topA.eps,height=0.22in}} \\
$(\overline{D}^0\rightarrow K^+\pi^-)$  &   &   &  &  &  \\
$(K^{*0}\rightarrow K^+\pi^-)$  &   &   &  &  &  \\
  &   &   &  &  &  \\
{\bf {\boldmath $B^0\rightarrow D^0 K^{*0}$}}
	& 0.0077  & $<3$ & high-$p^{}_T$  & 2/5 & 
		\mbox{\epsfig{file=topA.eps,height=0.22in}} \\
$(D^0\rightarrow K^-\pi^+)$  &   &   &  &  &  \\
$(K^{*0}\rightarrow K^+\pi^-)$  &   &   &  &  &  \\
  &   &   &  &  &  \\
{\bf {\boldmath $B^0\rightarrow D^0_{CP} K^{*0}$}}
	& 0.0035  & $<1.5$ & high-$p^{}_T$  & 2/5 & 
		\mbox{\epsfig{file=topA.eps,height=0.22in}} \\
$(D^0_{CP}\rightarrow K^+ K^-)$  &   &   &  &  &  \\
$(D^0_{CP}\rightarrow\pi^+\pi^-)$  &   &   &  &  &  \\
$(K^{*0}\rightarrow K^+\pi^-)$  &   &   &  &  &  \\
\hline
  &   &   &  &  &  \\
{\bf {\boldmath $B^0\rightarrow \pi^{\pm} K^{\mp}$}}
	& 1.4  & 260 & high-$p^{}_T$  & 1/2 & 
		\mbox{\epsfig{file=topG.eps,height=0.22in}} \\
  &   &   &  &  &  \\
{\bf {\boldmath $\overline{B}^0\rightarrow D^{*+}\pi^-$ }}
	& 7.4  & $<3100$  & high-$p^{}_T$  & 2/5  & 
		\mbox{\epsfig{file=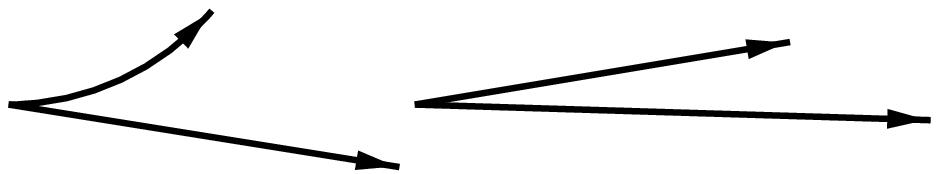,height=0.22in}} \\
$(D^{*+}\rightarrow D^0\pi^+)$  &   &   &  &  &  \\
$(D^0\rightarrow K^-\pi^+)$  &   &   &  &  &  \\
  &   &   &  &  &  \\
{\bf {\boldmath $B^0_s\rightarrow \rho^0 K^0_S$}}
	& 0.003  & $\ll 1.5$  & high-$p^{}_T$  & 2/5  & 
		\mbox{\epsfig{file=topA.eps,height=0.22in}} \\
  &   &   &  &  &  \\
\hline
\end{tabular}
}
\end{center}
\caption{Methods to measure the CKM angle $\gamma$ in \hb.
A ``$<$'' sign indicates that the trigger and reconstruction 
efficiencies assumed are optimistic.}
\label{tab:gamma}
\end{table}

\begin{table}[htb]
\begin{center}
\hbox{
\hspace*{-0.40in}
\begin{tabular}{|l|ccccc|}
\hline
{\bf Decay mode} &  {\bf BR} &  {\bf Rec.\,Evnts} & {\bf Trigger}  
& {\bf Vertices/}{\bf }  & {\bf Topology} \\
     &  {\bf ($\times 10^{-5}$)} & {\bf per year}  &    &
						{\bf Constraints}  &   \\
\hline
  &   &   &  &  &  \\
{\bf {\boldmath $B^0_s\rightarrow D^-_s\pi^+$}}
	& 12  & $<280$ & high-$p^{}_T$  & 1/5 & 
		\mbox{\epsfig{file=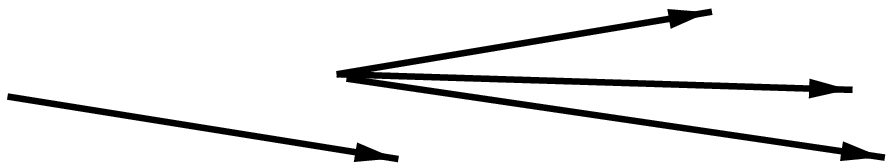,height=0.22in}} \\
$(D^-_s\rightarrow\phi\pi^-, K^{*0}K^-)$  &   &   &  &  &  \\
$(\phi\rightarrow K^+K^-)$        &   &   &  &  &  \\
$(K^{*0}\rightarrow K^+\pi^-)$    &   &   &  &  &  \\
{\bf {\boldmath $B^0_s\rightarrow D^-_s\pi^+\pi^-\pi^+$}}
	& 32  & $<740$ & high-$p^{}_T$  & 2/5 & 
		\mbox{\epsfig{file=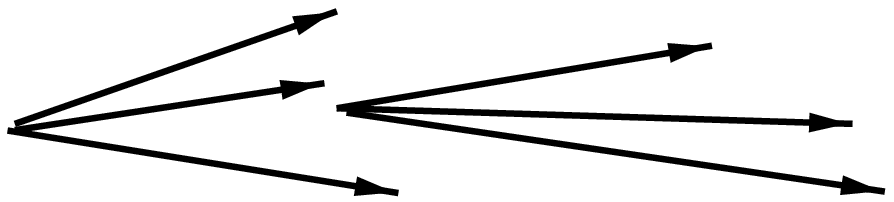,height=0.22in}} \\
  &   &   &  &  &  \\
{\bf {\boldmath $B^0_s\rightarrow D^-_s\ell^+\nu^{}_\ell$}}
	& 320  & 900 & high-$p^{}_T$ & 1/4 & 
		\mbox{\epsfig{file=topL.eps,height=0.22in}} \\
  &   &   &  &  &  \\
{\bf {\boldmath $B^0_s\rightarrow J/\psi\,\overline{K}^{\,*0}$}}
	& 0.52  & 100  & $\mu\mu/ee$  & 1/4  & 
		\mbox{\epsfig{file=topJ.eps,height=0.32in}} \\
$(J/\psi\rightarrow\ell^+\ell^-)$  &   &   &  &  &  \\
$(\overline{K}^{\,*0}\rightarrow K^-\pi^+)$  &   &   &  &  &  \\
\hline
  &   &   &  &  &  \\
{\bf {\boldmath $\overline{B}^0_s\rightarrow D^-_s K^+$}}
	& 0.32  & $<7.4$ & high-$p^{}_T$  & 1/5 & 
		\mbox{\epsfig{file=topL.eps,height=0.22in}} \\
  &   &   &  &  &  \\
{\bf {\boldmath $B^0_s\rightarrow D^-_s K^+$}}
	& 0.63  & $<15$ & high-$p^{}_T$  & 1/5 & 
		\mbox{\epsfig{file=topL.eps,height=0.22in}} \\
\hline
  &   &   &  &  &  \\
{\bf {\boldmath $B^0_s\rightarrow K^+ K^-$}}
	& 1.5  & $35$ & high-$p^{}_T$  & 1/2 & 
		\mbox{\epsfig{file=topG.eps,height=0.22in}} \\
\hline
  &   &   &  &  &  \\
{\bf {\boldmath $B^0_s\rightarrow J/\psi\phi$}}
	& 5.5  & $110$ & $\mu\mu/ee$  & 1/4 & 
		\mbox{\epsfig{file=topJ.eps,height=0.32in}} \\
$(J/\psi\rightarrow\ell^+\ell^-)$  &   &   &  &  &  \\
$(\phi\rightarrow K^+K^-)$  &   &   &  &  &  \\
  &   &   &  &  &  \\
\hline
\end{tabular}
}
\end{center}
\caption{$B^0_s$ decay modes to be collected in \hb. These modes 
can be used to measure $\Delta m^{}_s$ and $\Delta\Gamma^{}_s$, to 
constrain $\sin(2\phi^{}_M +\gamma)$, and to search for new physics. 
All $D^-_s$ final states are reconstructed via 
$D^-_s\rightarrow\phi\pi^-\rightarrow K^+K^-\pi^-$ or
$D^-_s\rightarrow K^{*0}K^-\rightarrow K^+\pi^-K^-$.
A ``$<$'' sign indicates that the trigger and reconstruction 
efficiencies assumed are optimistic.}
\label{tab:bsphysics}
\end{table}

\begin{table}[htb]
\begin{center}
\begin{tabular}{|l|ccc|}
\hline
   &  {\bf {\boldmath $D^{(*)}_s\ell^+\nu$}}
  &  {\bf {\boldmath $D^-_s\pi^+$}}
  &  {\bf {\boldmath $J/\psi\,\overline{K}^{\,*0}$}}  \\
  &  & {\bf {\boldmath $D^-_s\pi^+\pi^-\pi^+$}} &   \\
\hline
{\boldmath $n$} (events/yr) & 
		$\sim$\,900 & $\sim$\,400 & $\sim$\,100  \\
{\boldmath $f^{}_s$} (signal fraction) & 0.35 & 0.6  & 0.9  \\
{\boldmath $\eta^{}_{\rm mistag}$} (prob.\ of mistag) & 
		0.24  & 0.33  & 0.33  \\
{\boldmath $\sigma^{}_z$} (vertex resolution) & 
		600~$\mu$m & 450~$\mu$m & 300~$\mu$m  \\
{\boldmath $\sigma^{}_t$} (proper time  resolution) & 
		0.11~ps & 0.081~ps & 0.054~ps  \\
\hspace*{0.20in} ($\langle p^{}_B\rangle\approx 100$~GeV/$c$) &  &  &   \\
\hline
{\boldmath $\Delta m^{}_s$ reach in one year}  & & & \\
\hspace*{0.20in} $(S/N = 1.645)$  & 
		12~ps$^{-1}$ & 13~ps$^{-1}$ & 14~ps$^{-1}$  \\
{\boldmath $\Delta m^{}_s$ reach in two years}  &
		   14   &   17   &   21   \\
{\boldmath $\Delta m^{}_s$ reach in three years}  &
		   15   &   18   &   24   \\ 
   &   &     &   \\
 & \multicolumn{3}{c|}{(World limit: $\Delta m^{}_s>14.3$ @ 95\% C.L.)}   \\
\hline
\end{tabular}
\end{center}
\caption{Decay modes to be used to measure or constrain the mass 
difference $\Delta m^{}_s$, and the lower limits obtained in one, 
two, and three years of running.}
\label{tab:bsmixing}
\end{table}

\end{document}